\documentclass[prb,reprint,floatfix]{revtex4-1} 

\newlength{\figwidth}
\usepackage[utf8]{inputenc}
\usepackage{amsmath}  
\usepackage{amsfonts} 
\usepackage{graphicx} 
\usepackage{color}
\usepackage{siunitx}
\usepackage{hyperref}
\usepackage{epstopdf}

\newcommand{\LP}{LoggerPro\textsuperscript{\textregistered}}

\begin{document}


\title{Low-cost ultrasonic distance measurement in a mechanical resonance experiment}

\author{William D. Joysey}
\email{joyse1wd@cmich.edu} 
\author{Axel Mellinger}
\email{axel.mellinger@cmich.edu}
\affiliation{Department of Physics, Central Michigan University, Mount Pleasant, MI 48859}



\date{\today}

\begin{abstract}
We present a low-cost, dual-probe position sensor in a mechanical resonance experiment suitable for deployment in large lab courses with multiple stations. The motion of the two ends of a driven, damped spring oscillator is recorded with US-100 ultrasonic distance sensors and ESP8266 microcontrollers. Sensor lag is compensated via a modified Savitzky-Golay filter. Data is downloaded to a computer via Wi-Fi in a format suitable for analysis in Logger Pro\textsuperscript{\textregistered}. Due to the simple and fast data acquisition process, students can gather sufficient data to plot curves of the amplitude and phase lag as a function of driving frequency.  
\end{abstract}

\maketitle 

\section{Introduction}
\begin{figure}[tbp]
\centering
\includegraphics[width=\figwidth]{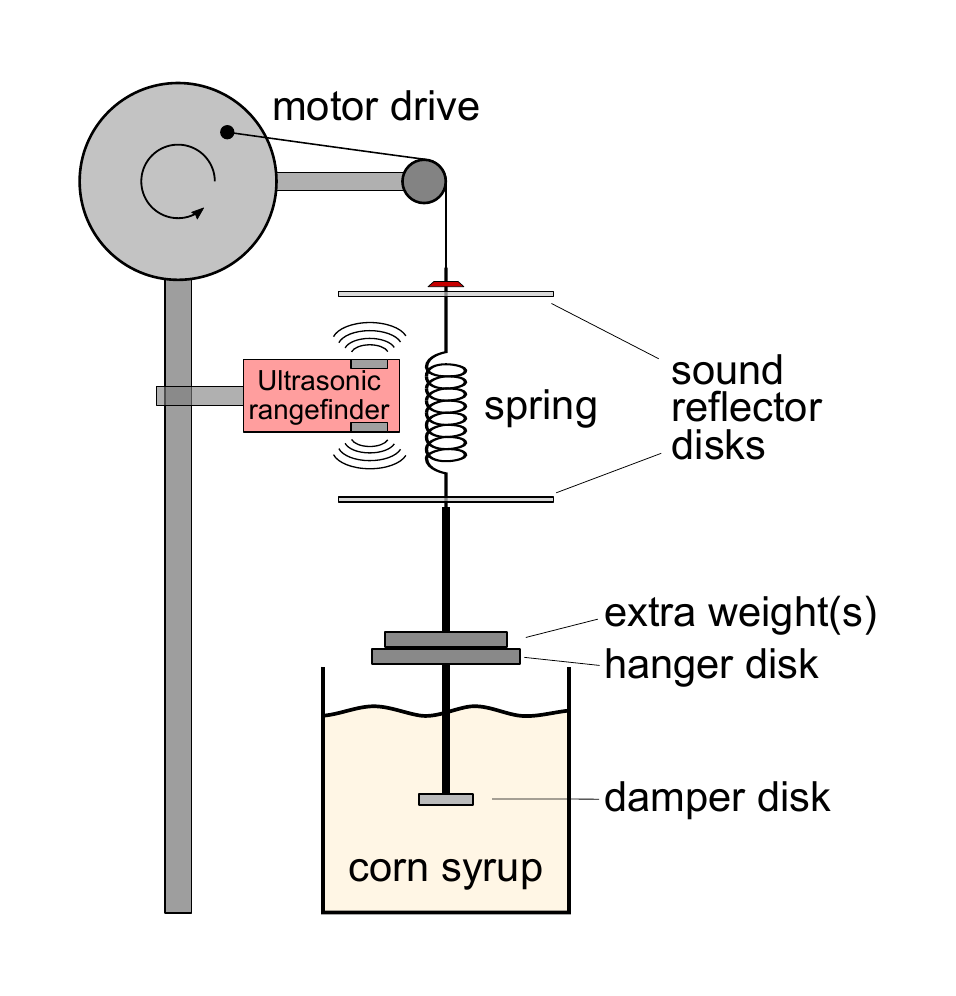}
\caption{Schematic view of the apparatus for apparatus for studying driven oscillations.}
\label{fig:apparatus}
\end{figure}

Mechanical resonance is a key concept in physics and engineering\cite{Bauer-Westfall}, with many practical applications such as automotive vehicle suspensions and musical instruments. Resonance also occurs in electrical circuits, optics and at the atomic and nuclear scale.
Therefore, studying driven, damped harmonic motion is a desirable experiment in many undergraduate physics lab courses. However, there appears to be a dearth of affordable, commercial equipment for such experiments. In some European countries, Pohl's pendulum (a spring-loaded driven disk with electromagnetic induction damping) is a popular experiment in first-year lab courses\cite{Pohl,Pohl-PHYWE}, but it appears to be less common elsewhere. Moreover, the cost of such an apparatus is substantial, especially for large introductory physics lab courses with 10 or 20 identical stations. In this paper, we discuss a simple motor-driven coil spring oscillator (Fig.~\ref{fig:apparatus}) with ultrasonic distance measurement.

The effect of mechanical resonance in a mass-spring oscillator is two-fold: (a) an increase in amplitude and (b) a phase lag between the free and driven ends of the spring oscillator. While the amplitude of a low-frequency oscillation can be measured with a simple ruler or meter stick, an accurate phase shift measurement requires that the positions of both the driven and free end of the spring be recorded as a function of time. 

There is a variety of methods for measuring time-dependent positions in a physics lab experiment, including the venerable spark tape\cite{Ols1992}, video analysis\cite{VideoAnalysis}, photogates\cite{Gal2013} and ultrasonic rangefinding\cite{Gat1992}.
In recent years, a number of compact, very low cost ultrasonic rangefinders have become popular in the maker community, and have also found use in physics labs to study simple harmonic \cite{Gal2014,Gon2017} and free-fall\cite{Moy2018} motion.
Their output can be recorded and processed with microcontrollers, such as the popular Arduino family,\cite{Bou2017,Lav2018} or the Wi-Fi-enabled system-on-a-chip ESP8266 and ESP32 microcontrollers.\cite{Ben2018} Ultrasonic rangefinding is non-contacting (i.\,e.\ does not introduce extra friction), significantly less tedious than video analysis, has a resolution of \SIrange{1}{3}{mm} and is robust enough to provide reliable data in a teaching lab.

\section{Theory}
The theory of driven (forced) harmonic oscillators is covered by most introductory physics textbooks\cite{Bauer-Westfall}. We assume a mass $m$ on a spring with spring constant $k$, a damping force of $-bv$ (where $v=\dot{x}$ is the velocity and $b$ is the damping parameter) and a periodic driving force $F_0\cos(\omega_{\text{d}} t) = k A_\text{d} \cos(\omega_{\text{d}} t)$. The differential equation of motion is
\begin{equation}
m\ddot{x} +b\dot{x} + kx = k A_\text{d}\cos(\omega_{\text{d}} t + \phi_\text{d})\:.
\label{eq:motion_driven_oscillator}
\end{equation}
with the steady-state solution
\begin{equation}
x(t) = A \cos(\omega_{\text{d}} t + \phi) + x_0\:.
\label{eq:solution_driven_oscillator}
\end{equation}
where the amplitude $A$ is a function of the driving (angular) frequency $\omega_\text{d}$:
\begin{equation}
A(\omega_{\text{d}}) = \frac{A_\text{d} \omega_0^2}{\sqrt{\left(\omega_{\text{d}}^2-\omega_0^2\right)^2 + b^2 \omega_{\text{d}}^2/m^2}}\:,
\label{eq:A_vs_omega}
\end{equation}
Here, $\omega_0=\sqrt{k/m}$ is the natural angular frequency of the undamped oscillator. Due to inertia, the oscillation of mass $m$ lags the oscillation of the spring's driven end by a phase angle
\begin{equation}
\Delta\phi(\omega_{\text{d}}) = \phi_\text{d} - \phi(\omega_{\text{d}}) = \arctan\left( \frac{b\omega_{\text{d}}/m}{\omega_0^2-\omega_{\text{d}}^2} \right)
\label{eq:phi_vs_omega}
\end{equation}
With the experimental setup presented in this paper, both $A(\omega_{\text{d}})$ and $\Delta\phi(\omega_{\text{d}})$ can be measured.

\section{Apparatus}
\subsection{Mechanical Setup}
A schematic view of the apparatus is shown in Fig.~\ref{fig:apparatus}. A Pasco ME-8750 Mechanical Oscillator/Driver generates a low-frequency sinusoidal motion. A nylon thread connects the driver unit to a coil spring with a spring constant of approximately $k=\SI{20}{N/m}$. The bottom end of the spring is connect to a hanger with a 100 or \SI{200}{g} mass and a damper disk (\SI{25}{mm} diameter) that oscillates in a beaker filled with corn syrup. The corn syrup was slightly diluted with water to obtain a damping constant of about \SI{0.6}{kg/s}. The voltage applied to the motor (\SIrange{0}{12}{V}) controls the oscillation frequency ($f_\text{d}\approx\,$\SIrange{0}{2.5}{Hz}).

The dual ultrasonic position sensor is mounted alongside the spring. It ultrasound pulses are reflected by two acrylic disks, mounted at the top and bottom end of the spring, respectively.

\subsection{Electrical and Software Setup}
Distances are measured by two US-100 ultrasonic rangefinders, widely available on Ebay and from robotic equipment vendors  (cost approximately US\,\$4 per item). 
Unlike the older (and slightly cheaper) HC-SR04 and US-015 sensors\cite{Gal2014}, the US-100 can operate in two modes: 
\begin{itemize}
\item the traditional trigger/echo mode, where a trigger signal of at least \SI{10}{\micro\second} will cause the emission of a train of short ultrasound pulses. The sensor then generates a \SI{5}{V} pulse on its echo pin, with a length equal to the ultrasound round-trip time.
\item a serial communication protocol (9600 baud, 8 bits, no parity), where the distance measurement is initiated by sending a \texttt{0x55} character to the sensor's trigger pin, and the distance to the object in mm is returned as a 2-byte integer on the echo pin.
\end{itemize}
For improved accuracy, the US-100 uses an on-board thermometer to calculate the speed of sound as a function of ambient temperature (serial mode only).
Due to its compact size (approx.\ $\SI{40}{mm}\times\SI{15}{mm}$) and its short minimum detection distance of \SI{2}{cm}, the US-100 rangefinder can be used in tight spaces.

In our setup, the rangefinders measure the distance to plastic disks attached to the top and bottom end of the spring (Fig.~\ref{fig:apparatus}). Since they are located \textit{between} the disks, the sign of the position reported by the top sensor needs to be inverted. In addition, an \SI{18}{cm} offset is applied, so that both positions are reported as positive values. 
The data from the distance sensors is collected using a WeMos D1 mini Wi-Fi microcontroller (cost approx.\ US\,\$5). 
Status messages are displayed on a 0.96-inch organic LED (OLED) display (cost US\,\$5). 

\begin{figure}[tbp]
\centering
\includegraphics[width=\figwidth]{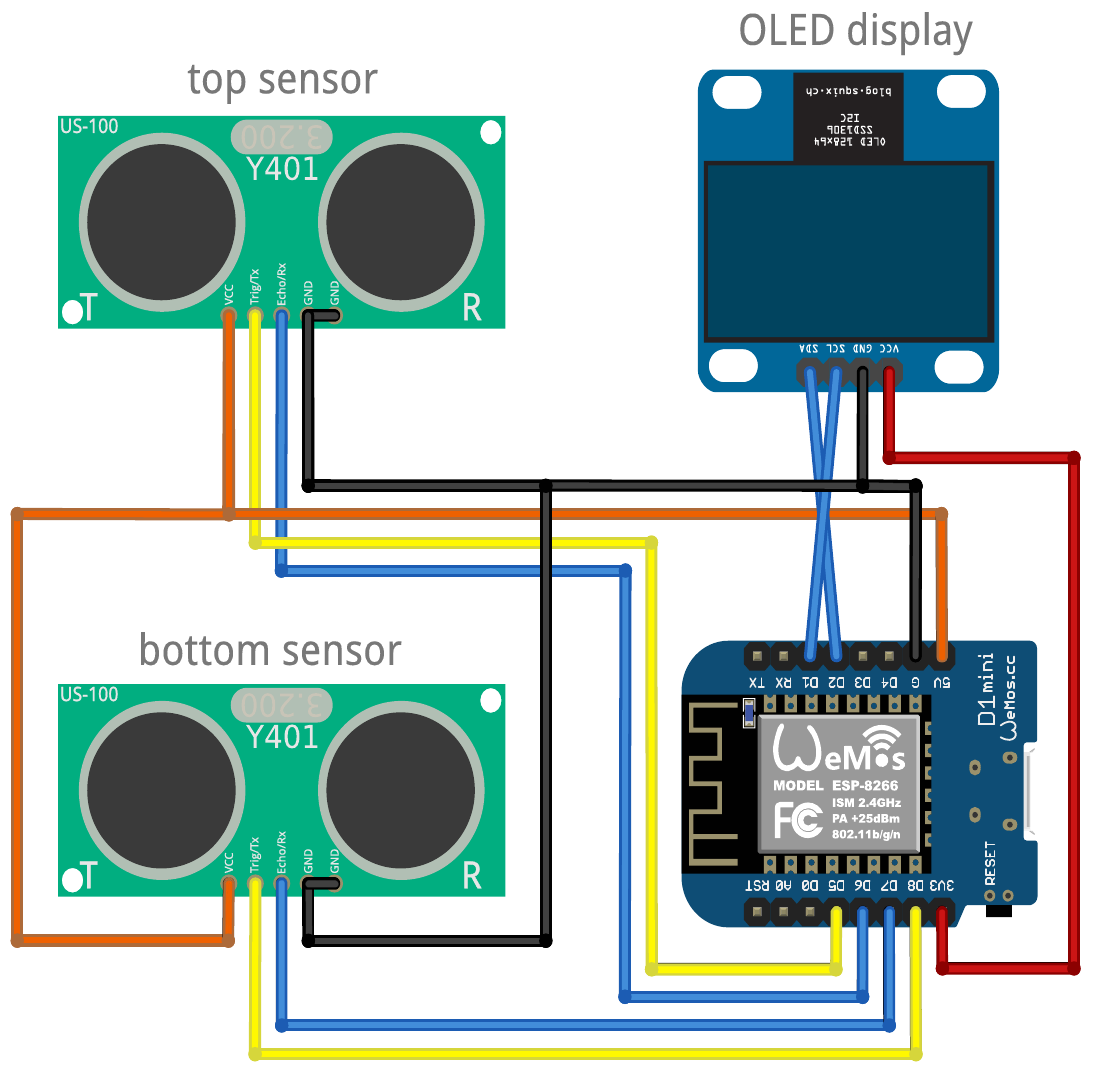}
\caption{Wiring schematics of the ESP8266-based WeMos D1 microcontroller, the US-100 ultrasonic distance sensors and the OLED display.}
\label{fig:schematics}
\end{figure}

\begin{figure}[tbp]
\centering
\includegraphics[width=\figwidth]{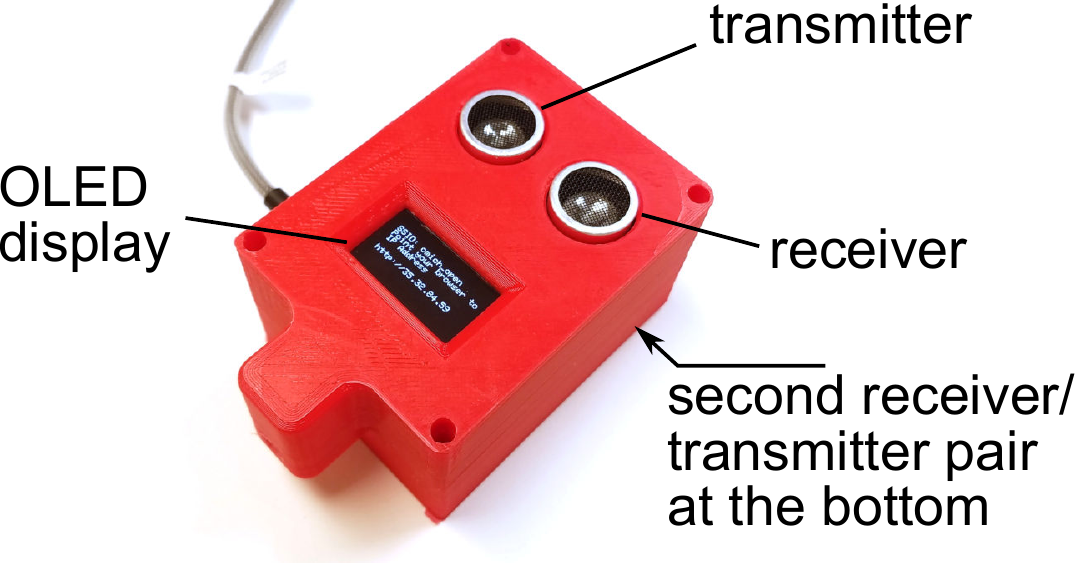}
\caption{Ultrasonic sensor head with its 3D-printed enclosure.}
\label{fig:sensorbox}
\end{figure}

The components are wired as shown in Fig.~\ref{fig:schematics} and mounted in a 3D-printed enclosure (Fig.~\ref{fig:sensorbox}). Power is supplied via a micro-USB cable, either from the USB port of a computer, or a \SI{5}{V} USB power supply. The ESP8266 software was written in C++ using the Arduino IDE\cite{Arduino} and is available for download on Github.\cite{GitHub} The internal web server uses the jQuery and Chart.js JavaScript libraries. 

\begin{figure}
    \centering
    \includegraphics[width=\figwidth]{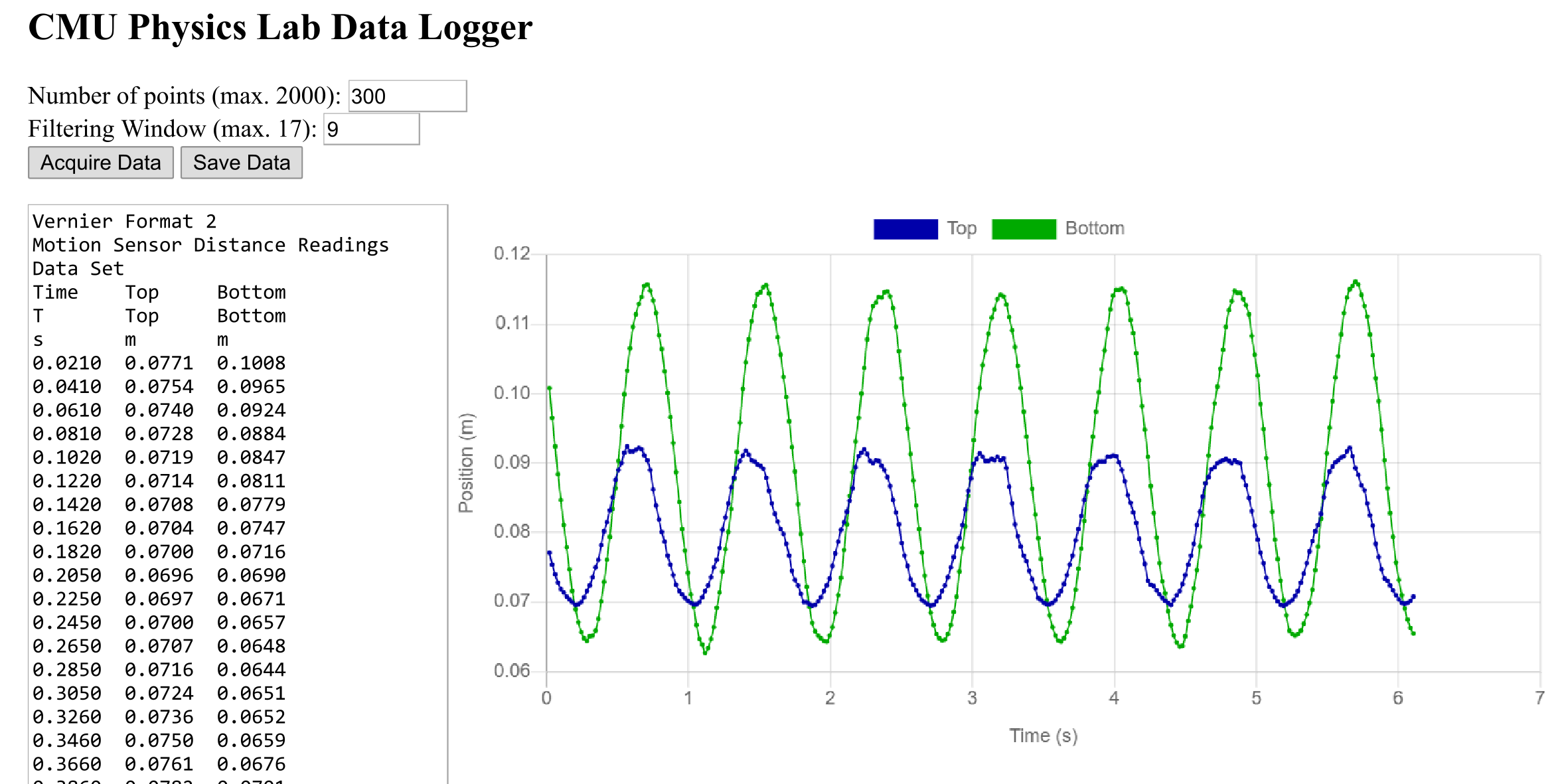}
    \caption{Web interface of the ESP8266 microcontroller, showing the table output (left) and the graph of position vs.\ time (right).}
    \label{fig:form}
\end{figure}

When the unit is powered up, a greeting message is shown on the OLED display. The ESP8266 attempts to connect the university's ``open'' (not password protected) Wi-Fi network. Should this network be unavailable, the ESP8266 switches to ``soft access point'' mode, setting up its own Wi-Fi network. Next, the a message similar to\\[2ex]
\parbox[t]{\figwidth}{%
\texttt{SSID: cmich\_open\\ 
Point your browser to IP Address http://aa.bb.cc.dd}}\\[2ex]
is displayed. The student then navigates to the indicated URL using a web browser. Initially, a simple web form is shown  (Fig.~\ref{fig:form}). The only user-defined input parameters are the number of data points to be recorded (10-2000), and the window width of the Savitzky-Golay (S-G) smoothing filter.\cite{Sav1964}

A measurement is initiated by clicking the ``Acquire Data'' button. Upon completion, a graph of the positions (in meters) versus time (in seconds) is displayed. In addition, the data is returned as a 3-column ASCII table, which can be saved to the computer's hard drive with the ``Save Data'' button. The table includes a few header lines to make the file recognizable by Vernier's \LP\ software.\cite{Vernier}

\section{Results and Discussion}

\subsection{US-100 Linearity Test}
\begin{figure}
    \centering
    \includegraphics[width=\figwidth]{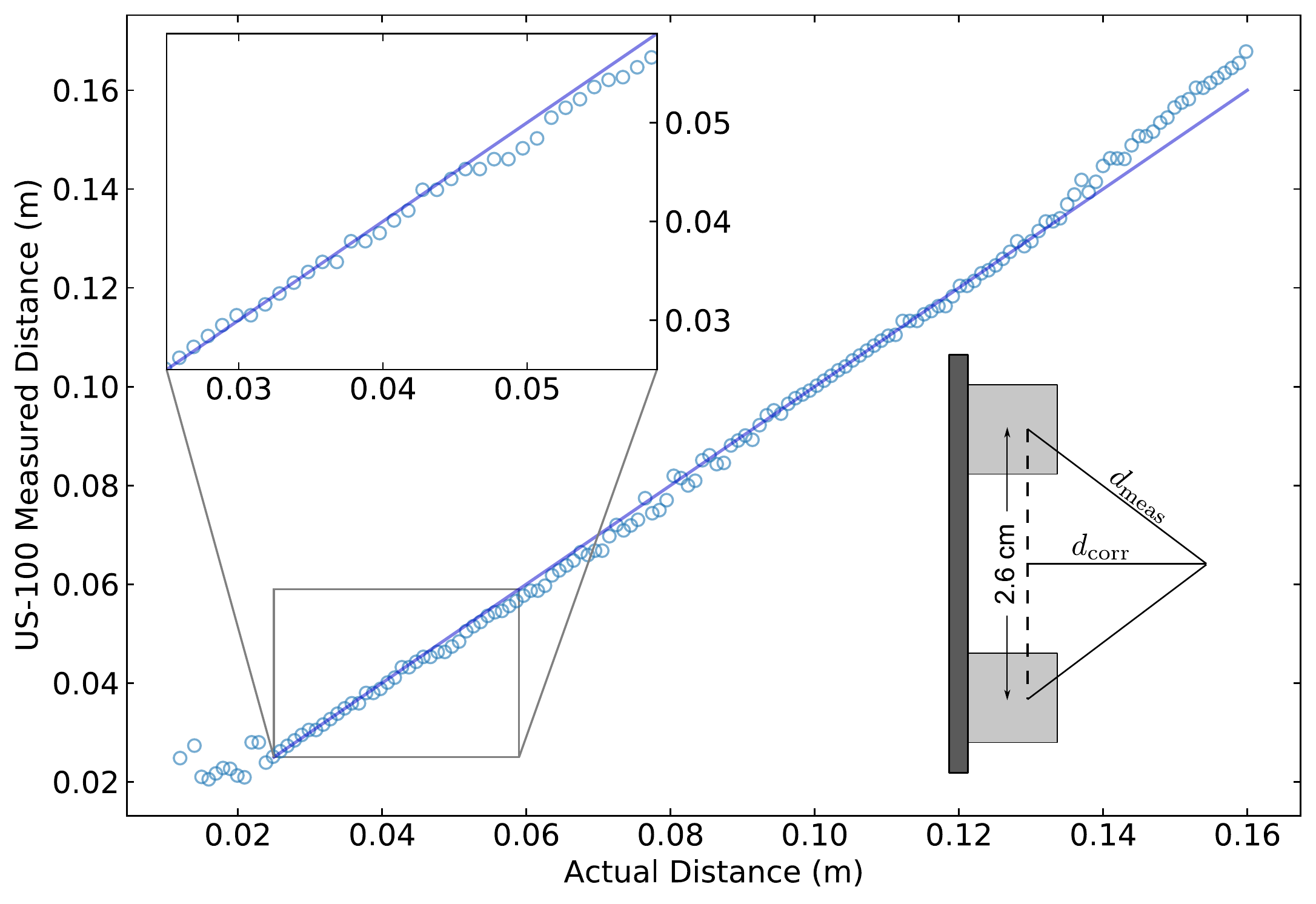}
    \caption{US-100 measurements vs.\ actual distance.  The solid line is the identity function $F(x)= x$. The bottom right inset shows the parallax correction.}
    \label{fig:Linearity}
\end{figure}

    To verify the accuracy of US-100 ultrasonic rangefinder it was mounted to a Zaber T-LSR 300B motorized linear slide and moved relative to a fixed target. Fig.~\ref{fig:Linearity} shows good linearity for distances from \SIrange{2.5}{13}{cm}. At \SIrange{8}{10}{cm} the sensor exhibits some non-monotonic behavior, which is a known property of low-cost ultrasonic rangefinders\cite{Pilling}. Valid data is returned for distances as low as \SI{2.5}{cm}. At small distances the finite distance $b=\SI{2.6}{cm}$ between transmitter and receiver has to be taken into account\cite{Gal2014}, as shown in the inset of Fig.~\ref{fig:Linearity}. Using the Pythagorean theorem, the corrected distance is $d_\text{corr} = \sqrt{d_\text{meas}^2 - (b/2)^2}$.

\subsection{Driven Oscillator}
\subsubsection{Amplitude}
\begin{figure}[tbp]
\centering
\includegraphics[width=\figwidth]{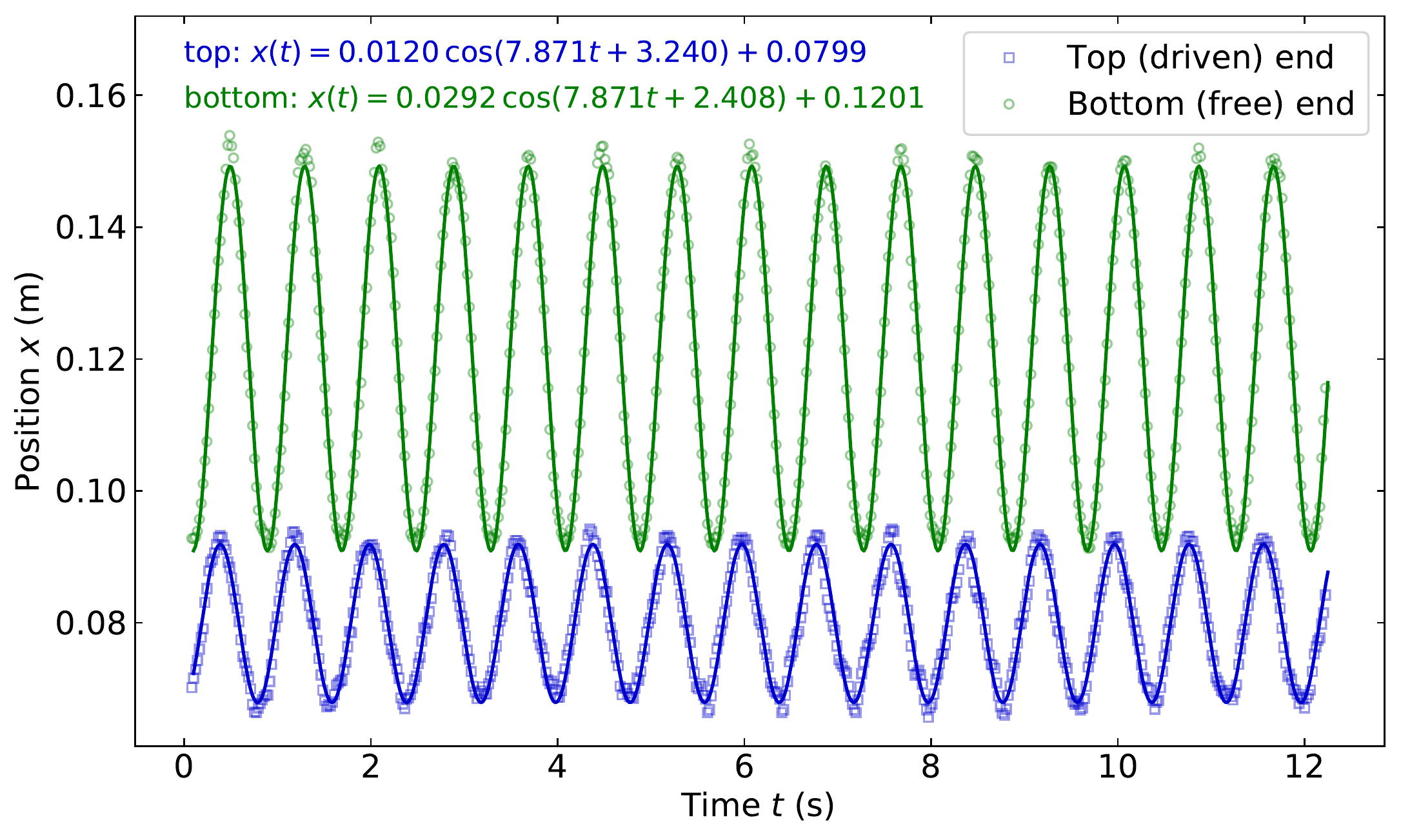}
\caption{Measured positions of the top (driven) and bottom (free) end of the spring. The solid lines are least square fits based on Eq.~(\ref{eq:fitfunc}). From the fit parameters (shown in SI units at the top of the graph), we obtain an amplitude ratio $A/A_\text{d}=2.43$, a phase lag  $\Delta\phi=\SI{0.832}{rad}$, and an angular frequency $\omega_{\text{d}}=\SI{7.871}{\per\second}$.}
\label{fig:measurement}
\end{figure}

Fig.~\ref{fig:measurement} shows a typical measurement result for a driving frequency $\omega_\text{d}$ slightly below resonance. The increase in amplitude at the bottom end of the spring is clearly visible, as is the phase lag.

Once the data has been imported into  \LP, the students determine the amplitude and phase by fitting functions of the type
\begin{equation}
x(t) = A\cos(Bt+C)+D
\label{eq:fitfunc}
\end{equation}
to the top and bottom distances, where $A$, $B$, $C$ and $D$ are fit parameters. 
Sometimes, the fit fails and returns a nearly straight line (i.\,e.\ a cosine curve with a very small amplitude). In that case, students are encouraged to manually optimize the initial fit parameters before trying the automatic curve fit again.

\begin{figure}[tbp]
\centering
\includegraphics[width=\figwidth]{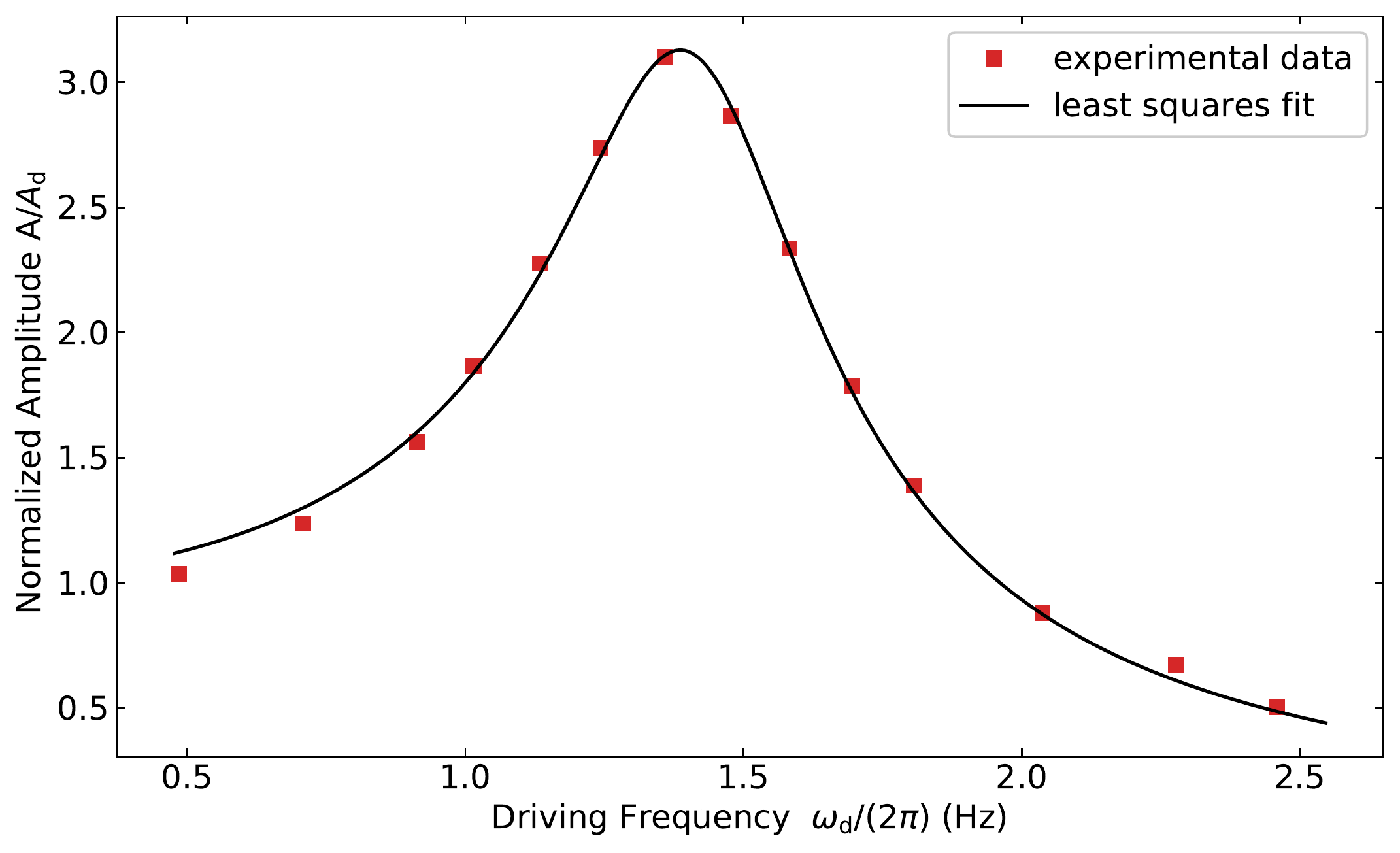}
\caption{Amplitude ratio $A(\omega)/A_\text{d}$ vs.\ driving frequency. The amplitudes were determined by fitting Eq.~(\ref{eq:fitfunc}) to the experimental data, as shown in Fig.~\ref{fig:measurement}. The solid curve is a least squares fit of Eq.~(\ref{eq:A_vs_omega}) to the experimental amplitude ratios.}
\label{fig:amplitude}
\end{figure}

From the least squares fit parameters of Eq.~(\ref{eq:fitfunc}), the amplitude ratio $A(\omega)/A_\text{d}$ of the free end over the driven end can be calculated and plotted as a function of frequency (Fig.~\ref{fig:amplitude}). The results are in very good agreement with the curve of Eq.~(\ref{eq:A_vs_omega}).

\subsubsection{Phase-shift correction}
\begin{figure}[tbp]
\centering
\includegraphics[width=\figwidth]{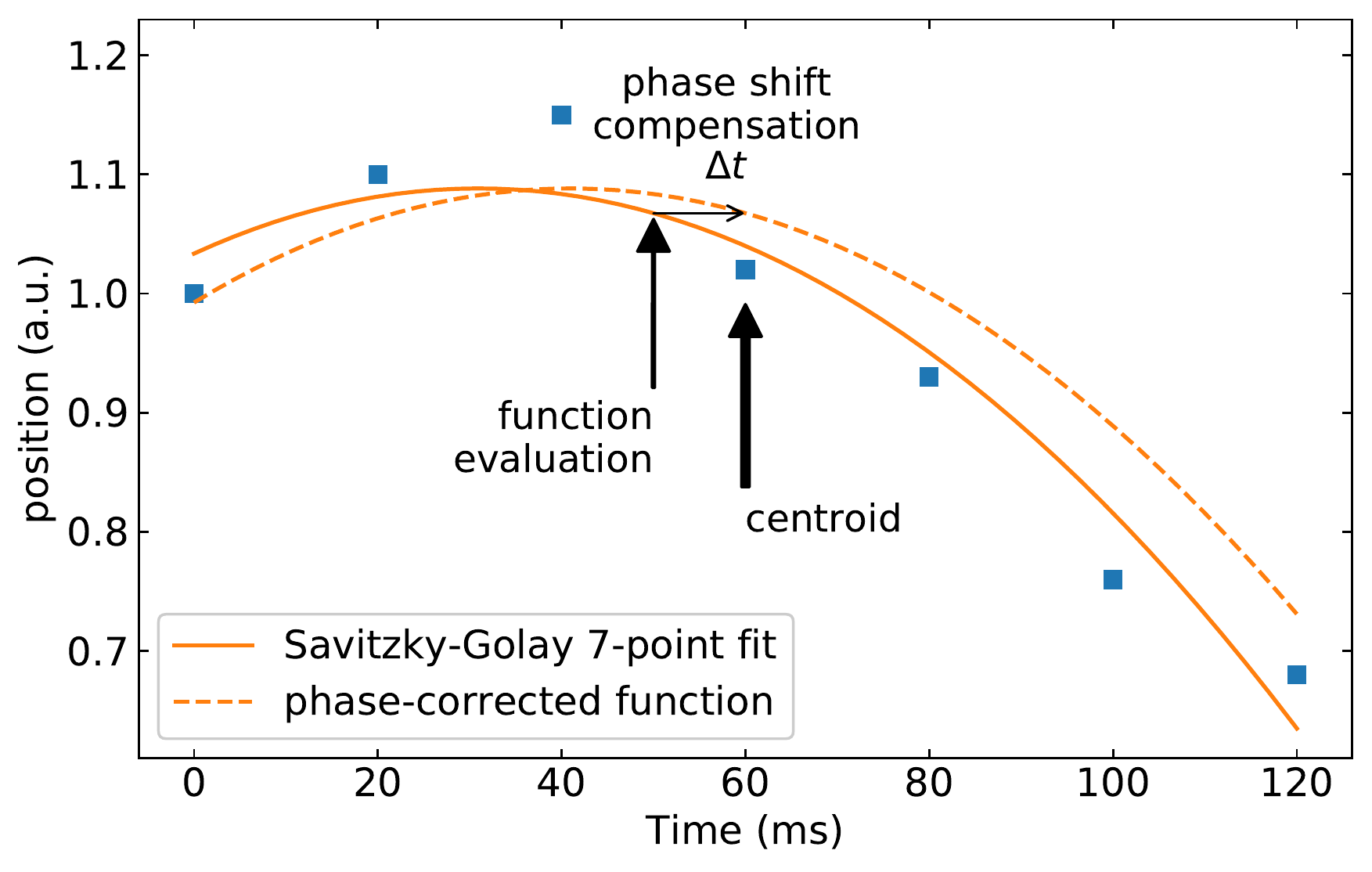}
\caption{Example of phase-shift correction for a 7-point quadratic Savitzky-Golay filter. To correct for the sensor lag $\Delta t$, the fitted polynomial is evaluated at a time $\Delta t$ \textit{earlier} than the centroid.}
\label{fig:SG_Filter}
\end{figure}

The two position measurements are taken approximately $\Delta t = \SI{10}{ms}$ apart, due to the time required by the US-100 rangefinder to acquire and transmit a distance reading. At the shortest oscillator period (about \SI{0.4}{s}), this equals a phase difference of \SI{0.16}{rad}. While not a large amount, it becomes a noticeable artifact in a plot of phase lag versus frequency. It is, of course, possible to record separate time stamps for each rangefinder. However, handling two time columns introduces an additional workload for the students, and may prevent them from plotting both positions in a single graph.

Instead, we compensate this phase shift through a modification of the S-G filter. In the example shown in Fig.~\ref{fig:SG_Filter}, a polynomial of order $n=2$ is fitted to a moving window of $w=7$ data points. Normally, this polynomial is evaluated at the centroid (here: $t_\text{c} = \SI{60}{ms}$). To compensate the delay of the second (bottom) sensor, we instead evaluate the polynomial for the bottom data at a time \SI{10}{ms} \textit{earlier} than the centroid. This is easily done using the first and second derivatives provided by the S-G filter: 
\begin{equation}
x_\text{corr} = x_\text{c} - \left.\frac{dx}{dt}\right|_{t_\text{c}} \Delta t + \frac{1}{2} \left.\frac{d^2x}{dt^2}\right|_{t_\text{c}} (\Delta t)^2
\end{equation}

\begin{figure}[tbp]
\centering
\includegraphics[width=\figwidth]{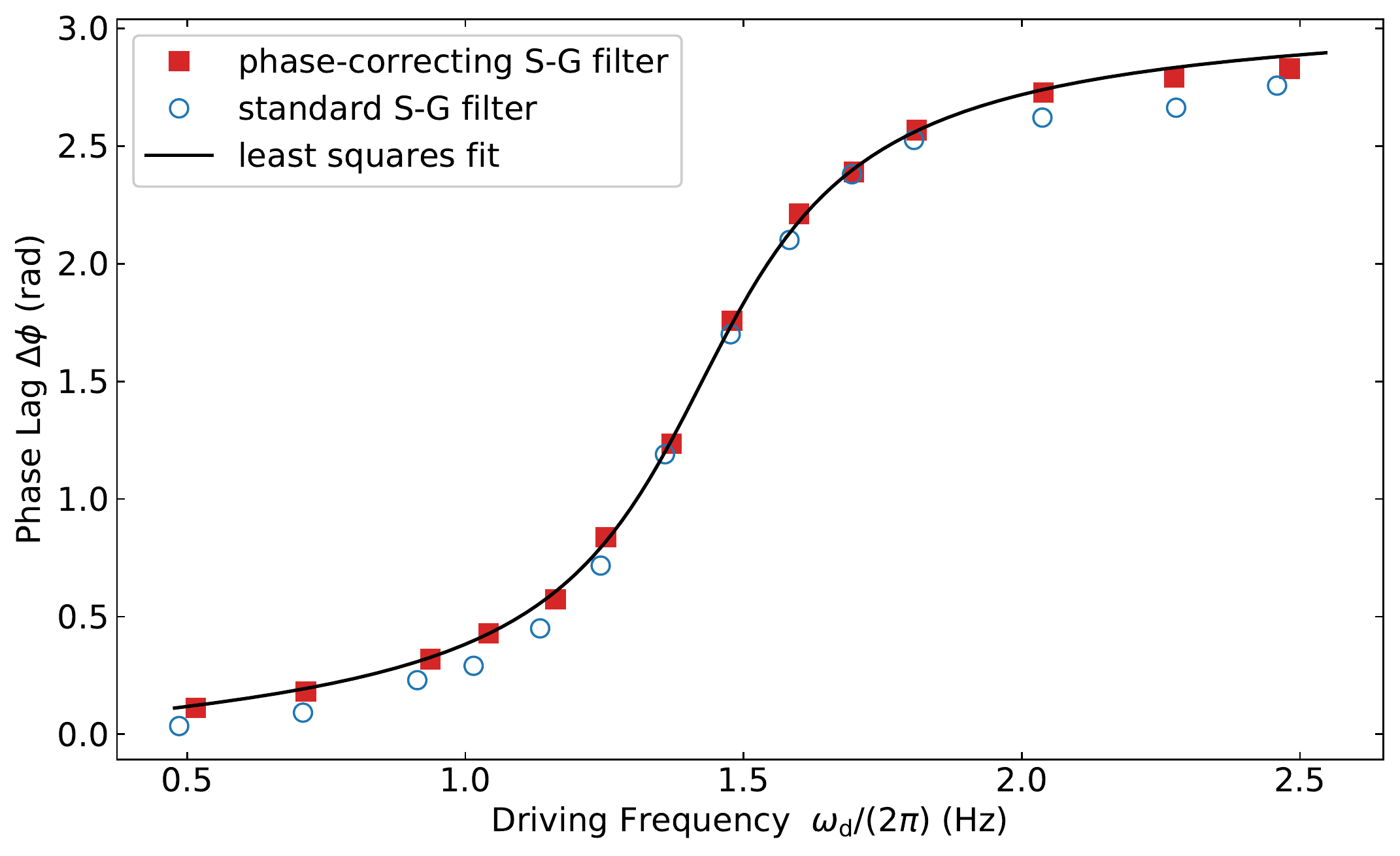}
\caption{Phase lag vs.\ driving frequency, with and without the phase-shift correcting S-G filter (see text). The solid curve is a least squares fit of Eq.~(\ref{eq:phi_vs_omega}) to the phase-corrected data.}
\label{fig:phase}
\end{figure}

The phase shift correction is built into the ESP8266 code and is normally not revealed to students in introductory lab courses (although it could become a discussion point in an advanced physics lab). The S-G filter coefficients were calculated using the equations given by Gorry\cite{Gor1990}, which correctly handle the first and last points in the data set, for which the filter kernel is non-symmetric (the number of points to the left and right differs). The effectiveness of the phase-shift correction can be seen in the phase lag plot of Fig.~\ref{fig:phase}.

\section{Conclusion}
We designed and built a dual-head ultrasonic position sensor for a driven oscillator experiment. The material cost for the sensor is under US\$ 50. Data can be transferred via Wi-Fi to any device that has a web browser; no costly sensor interface is required. The output is in an ASCII text format compatible with \LP; other output formats can be easily implemented. The work flow of acquiring the $x(t)$ data, importing them into \LP and calculating amplitude, phase lag and frequency via a least squares fit is sufficiently fast for the students to record 12-15 frequencies within a 2-hour lab session, and to draw graphs of $A(\omega_\text{d})$ and $\Delta\phi(\omega_\text{d})$.

Compared to earlier uses of ultrasonic rangefinding in mechanical resonance experiments\cite{Gal2014,Gon2017}, the dual-head rangefinder allows the measurement of the phase lag. A modification to the Savitzky-Golay smoothing filter compensates for the \SI{10}{ms} delay introduced by sequentially reading out the rangefinders. This resulted in a noticeable improvement of the phase lag plot.

The ultrasonic rangefinder is suitable for use in a variety of other linear motion experiments, such as uniform and accelerated motion on low-friction air tracks. The sampling rate (50 samples/s for dual head recording, 100 samples/s when used with a single rangefinder) is fast enough to provide the students with a finely spaced data set.

%
%

\begin{acknowledgments}
We gratefully acknowledge the help of Ray Clark and Mark Wilson in designing and building the resonance experiment and the sensor head. A.M. thanks Andrei Neacsu for many stimulating discussions on microcontrollers and sensors.
\end{acknowledgments}

\bibliographystyle{aipnum4-1}
\bibliography{Joysey+Mellinger_Ultrasound_Distance_Monitor}

\end{document}